\documentclass[sigconf]{acmart}
\usepackage{amsmath,amsfonts}
\usepackage{algorithmic}
\usepackage{graphicx}
\usepackage{textcomp}
\usepackage{xcolor}
\usepackage{wrapfig}
\usepackage{color,soul}
\usepackage{multirow}
\usepackage{url}
\usepackage{tikz}
\usepackage{comment}
\usepackage{subcaption}
\usepackage{graphicx}
\usepackage[absolute]{textpos}
\usepackage[flushleft]{threeparttable}

\AtBeginDocument{%
  \providecommand\BibTeX{{%
    \normalfont B\kern-0.5em{\scshape i\kern-0.25em b}\kern-0.8em\TeX}}}

\copyrightyear{2022}
\acmYear{2022}

\begin{document}


\title{An Empirical Investigation of Worker Clusters in TopCoder}

\settopmatter{authorsperrow=4}

\author{Razieh Saremi} 
\email{rsaremi@stevens.edu}
\affiliation{%
  \institution{Stevens Institute of Technology}
  \city{Hoboken}
  \state{New Jersey}
  \postcode{07030}}

\author{Hamid Shamszare}
\email{hshamsza@stevens.edu}
\affiliation{%
  \institution{Stevens Institute of Technology}
  \city{Hoboken}
  \state{New Jersey}
  \postcode{07030}}
  
  \author{Marzieh Lotfalian Saremi}
\email{m.lotfa@encs.concordia.ca}
\affiliation{%
  \institution{Concordia University}
  \city{Montreal, QC}
  \state{Canada}
}
  
\author{Ye Yang}
\email{yyang4@stevens.edu}
\affiliation{%
  \institution{Stevens Institute of Technology}
  \city{Hoboken}
  \state{New Jersey}
  \postcode{07030}}

\renewcommand{\shortauthors}{Razieh Saremi, et al.}

\begin{abstract}
Software crowdsourcing platforms employ extrinsic rewards such as rating or ranking systems to motivate workers. Such rating systems are noisy and provide limited knowledge about workers' preferences and performance.  To develop better
understanding of worker reliability and trustworthiness in software crowdsourcing, this paper reports an empirical study conducted on more than one year’s real-world data from TopCoder, one of the leading software crowdsourcing platforms.

To do so, first, we create a bipartite network of active workers based on common task registrations. Then, we use  the Clauset-Newman-Moore graph clustering algorithm to identify worker clusters in the network. Finally, we conduct an empirical evaluation to measure and analyze workers' behavior per identified community in the platform by workers' rating. More specifically, workers' behavior is analyzed based on their performances in terms of reliability, trustworthiness, and success; their preferences in terms of efficiency and elasticity; and strategies in terms of comfort, confidence, and deceitfulness. 
The main result of this study identified four communities of active workers: mixed-ranked, high-ranked, mid-ranked, and low-ranked. This study shows that the low-ranked community associates with the highest reliable workers with an average reliability of 25\%, while the mixed-ranked community contains the most trustworthy workers with average trustworthiness of 16\%.
Such empirical evidences are beneficial to help
exploring resourcing options while understanding the relations among unknown resources to improve task  success.

\end{abstract}



\maketitle

\section{Introduction}

Crowdsourced software development (CSD) has gained increased popularity in recent years, however, there are many risks associated with it. 
Major risks are associated with the uncertainties in both the number  of registrants and quality of the received submissions from the unknown workers   \cite{eickhoff2013increasing}\cite{yang2015award}\cite{yu2015efficient}.
This raises the challenge of how to engage unknown workers and trust in their works \cite{ye2015crowd}. To encourage the engagement of crowdworkers, crowdsourcing platforms employ extrinsic rewards such as rating \cite{cheng2020building}.  To that end, crowdsourcing platforms employ different reputation systems to manage crowd rating based on their participation history. For example, the HITs rate is used in Amazon Mechanical Turk \cite{eickhoff2013increasing}, and a numeric developer rating based on the Elo rating algorithm is used in TopCoder \cite{saremi2017leveraging}.
However, such rating systems are noisy since the reliability or reputations of the workers are often unknown \cite{jagabathula2014reputation}, and only provide limited knowledge about workers’ preferences and performance. 
Also, according to the competitive exclusion principle\cite{hardin1960competitive}\cite{pocheville2015ecological}, rating systems can provide advantages for a group of workers over others in competition. The advantages can result in frequent wins of higher-rated workers in competitions. 


In general, competing for success over shared tasks among different rated workers creates a  complex network of workers.  
To understand the dynamic among workers in such network, it is essential to identify the existing communities in this network and investigate the behaviour of crowd workers per community. To that end, we need to develop a better understanding of the sensitivity in worker preference, performance, and engagement  \cite{kittur2008crowdsourcing}\cite{kittur2013future}\cite{yu2015efficient} per community. While there are available literature on investigating workers' behaviour and performances on different crowdsourced platforms \cite{karim2016decision}\cite{khanfor2017failure}\cite{saremi2017leveraging}\cite{ye2015crowd}, to the best of our knowledge, there is no investigation on the impact of workers 
behavior among different communities of crowd workers on workers' performance and success.

The objective of this study is to empirically investigate patterns and impacts of crowdworkers' performance and preferences in software crowdsourcing platforms in order to improve the success and efficiency of software crowdsourcing. In this study, first, we present a motivational example to shed light on the identified communities of active workers in software crowdsourcing platform of study, TopCoder. Then, we propose an empirical evaluation framework and develop an approach to characterize and analyze workers behaviour in each identified community. More specifically, this includes grouping similar rated workers and then studying the workers' preferences, performance, and strategy per community. The empirical study is conducted on more than one year’s real-world data from TopCoder\footnote{ \url{https://www.topcoder.com/}}, a leading software crowdsourcing platform with an online community of over 1.5M workers and 55k mini-tasks. The evaluation results show that: 1) there are four active communities of workers in the pool of workers: mixed-ranked, high-ranked, mid-ranked, and low-ranked; 2) low-ranked community provides the highest level of reliable workers to make a submission; 3) mixed-ranked community respects their expertise to compete fairly on a task; 4) mixed-ranked community applies strategies to assure their success.

The remainder of this paper is structured as follows. 
Section 2 presents background and related work. Section 3 outlines our research design. Section 4 presents the empirical results. Section 5 discusses the key findings of our study. Section 6 presents the conclusion and outlines a number of directions for future work.

\section{Background and Related Work}





 \subsection{Workers' Community in Crowdsourcing}
To encourage contribution and engagement, many communities including crowdsourcing adopt the use of extrinsic rewards such as rating as game elements for workers to compete in a non-gaming context \cite{cheng2020building}. Extrinsic rewards can increase the overall workers' engagement and commitment\cite{bista2012using}\cite{cavusoglu2015can}, motivation \cite{CHANDLER2013123}\cite{soliman2015understanding}\cite{stewart2009designing}, and collaboration \cite{gray2016crowd}; since they address a type of social need for some community members\cite{richter2015studying}.
In a CSD platform, a competitive environment not only influences the decisions of workers regarding which tasks to register and submit but also affects the way they react to their peers.
Therefore, many challenges in crowdsourcing such as worker motivation \cite{stewart2009designing}\cite{soliman2015understanding}\cite{CHANDLER2013123}, collaboration \cite{kittur2010crowdsourcing}, creativity \cite{kittur2013future}, performance, and trust \cite{laplante2016building}\cite{ye2015crowd} can be tackled from the online communities aspect \cite{ACAMPO2019775}. 
The principle of online communities has been used to improve the performance quality of crowd workers\cite{dow2012shepherding}. Also, it is reported that crowd workers tend to take collective actions to improve their own working situations\cite{salehi2015we}. Moreover, defining a community of workers in the pool of crowd workers is beneficial for both a platform and a task owner in different ways such as workers loyalty to the platform, collaboration among workers, and workers' trustworthiness \cite{ACAMPO2019775}.

\subsection{Workers' Performance in Crowdsourcing}

Software workers’ arrival to the platform and the pattern of taking tasks to complete are factors that shape the worker dynamic in a crowdsourcing platform, however, the reliability in returning the qualified tasks creates the dynamic of the platform. Generally, not only would the award associated with the task influence the workers’ interests in competitions\cite{yin2014monetary}, but also the number of registrants for the task, the number of submissions by individual workers, and the workers’ historical score rate would directly affect their final performance \cite{lakhani2010topcoder}\cite{saremi2017leveraging}. For newcomers or beginners, there is a time window required to improve and develop into an active worker \cite{faradani2011s}. Therefore, it is typical that the workers need to communicate with the task owner in order to better understand the problems to be solved \cite{kittur2013future}. Existing studies show that over time, registrants gain more experience, exhibit better performance, and consequently gain higher scores \cite{faradani2011s} \cite{archak2010learning} \cite{kittur2013future}. Still, there are workers who manage not only to win but also to raise their submission-to-win ratio \cite{difallah2016scheduling}. This motivates workers to develop behavioral strategies in TopCoder \cite{archak2010money} \cite{archak2010learning}. Moreover, the ranking mechanism used by TopCoder contributes to the efficiency of online competition and provides more freedom of choice for higher rate workers in terms of controlling competition level \cite{archak2010money}.

\subsection{Competition Strategies in Crowdsourcing }

Crowdsourcing a project inherently involves a concern for how reliable and trustworthy the unknown crowd workers are \cite{ye2015crowd}. It is reported that unreliable workers are not very interested in taking novel tasks that require creativity and abstract thinking \cite{eickhoff2013increasing}. 
Due to the diversity of workers with different individual skill levels, it is not practical for a requester to evaluate all the workers' trustworthiness \cite{ye2015crowd}, nor is there a clear record of workers' interaction in the pool of workers \cite{eickhoff2013increasing}. This fact creates a trust network among the worker community itself; which is a result of workers' rating, skill set, and task winning history. 
Such networks create opportunities for workers to apply different strategies to assure their success and increase their rank in the system.
One primary example is rank-boosting \cite{ipeirotis2010top} in Amazon Mechanical Turk, where workers mostly register for easy tasks or fake tasks that are uploaded by themselves in order to increase their rating or distort pursuit \cite{ye2015crowd}, in which workers quickly submit a possibly correct answer in order increase their benefits instead of working on the task and submitting an acceptable answer.
Another example is detecting cheap talk phenomena \cite{farrell1996cheap}\cite{archak2010money} in TopCoder. In CSD, higher-rated workers have more freedom of choice in comparison with lower-rated workers and can successfully affect the registration of lower-rated workers. To assure a softer and easier competition level, higher-rated workers register early for some specific projects while lower-rated workers must wait for higher-rated workers to make their choice. 

\section{Research Design}


Crowd workers are exchanging information and developing professional or social contacts. Such interaction leads to the creation of a \textit{complex network}, in which mutual motivation factors and personal preferences in taking a task among workers create different \textit{communities}. Workers in each community are densely connected to each other and loosely connected to the workers in the other communities in the network of workers \cite{malliaros2013clustering}.\\
In this study, to investigate dynamic behaviour patterns of crowd workers' in terms of performance, preference, and strategies in taking and submitting tasks based on different workers' communities, a preliminary analysis is conducted using data from TopCoder platform.


\subsection{Empirical Evaluation Framework}\label{model}

\begin{figure}[ht]
\centering
\includegraphics[width=0.8\columnwidth,keepaspectratio]{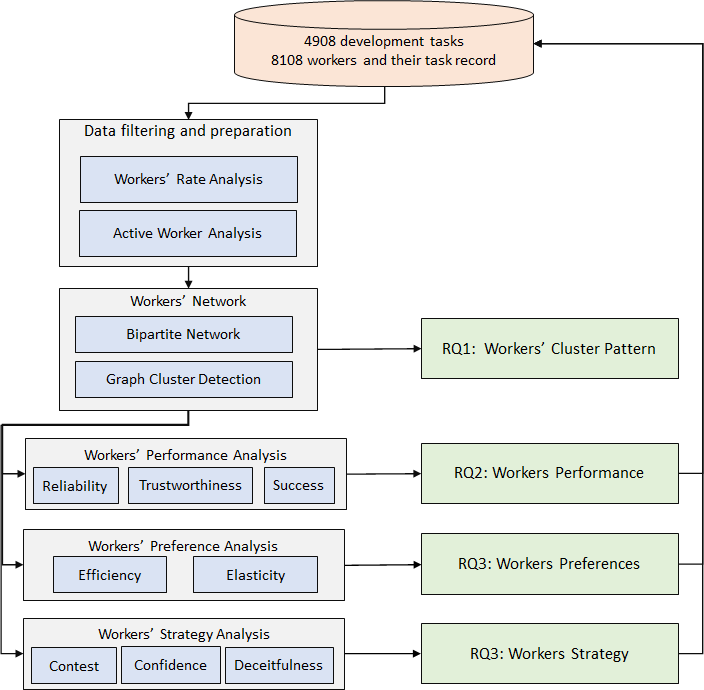}
\caption{Main Flow of the Proposed Empirical Evaluation Framework and Relationship to Research Questions}
\label{tam}
\end{figure}

To develop a better understanding of the workers' dynamics in different communities of workers in task supply and execution, we design four evaluation studies to investigate the differences in the preferences, performances, and strategies of workers in each identified community; and the effect of each community on delivering successful crowdsourcing tasks is studied.
The proposed evaluation framework is illustrated in figure \ref{tam}. Next, we explain the analysis steps and introduce a few working definitions used in this framework.

We created a bipartite network of active workers based on the number of common registered tasks among workers and detected the existing communities in the network. To detect the existing communities, we applied Clauset-Newman-Moore greedy modularity maximization algorithm \cite{PhysRevE.70.066111}, one of the most used community detection methods. This algorithm is hierarchical agglomeration therefore it supports graph classes and does not consider edge weights. The algorithm begins with each node in its own community and joins the pair of nodes that makes the most increase in modularity until no such pair exists. As a result, we identified four communities of workers in the network of active workers.
This analysis helps us to reveal the hidden relationships among the workers in the network.
Each community is a subset of the whole network of active workers \cite{girvan2002community}. Therefore, each identified community should contain a different type of workers in terms of expertise and rates. We followed the TopCoder rating belt, introduced in part \ref{data}, and grouped workers per community into 5 different belts of Gray, Green, Blue, Yellow, and Red. This helps us to investigate workers' behavior per identified community of workers. Since, in the entire dataset, we only had eight workers from Red belt (i.e 0.16\% of workers), this group of workers will be ignored in further analysis.
 

\textit{Identified Communities:} 
Figure \ref{network} shows the four identified communities in the dataset. 
Based on the diversity of workers' experience rate, table \ref{belts}, per identified community, the four communities are named as:

\begin{figure}[hbt!]
\centering
\includegraphics[width=0.8\columnwidth,keepaspectratio]{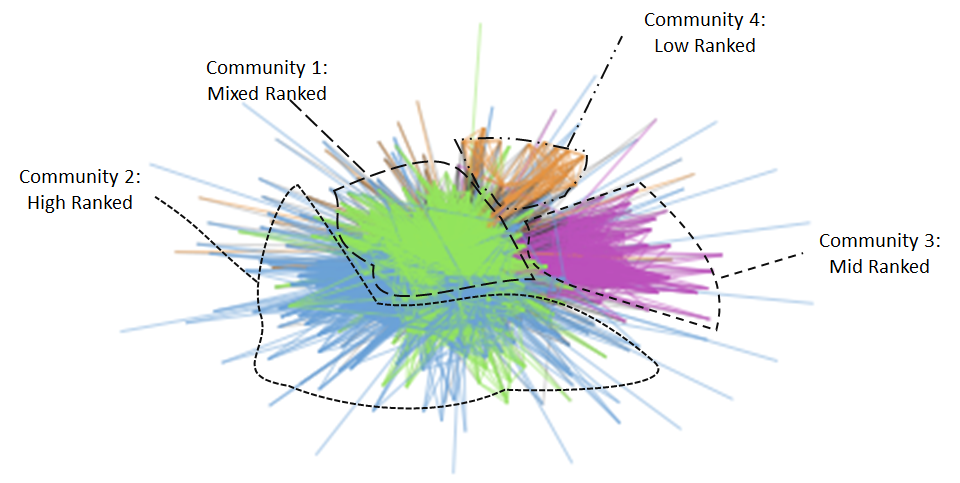}
\caption{Identified Communities in the Network of Active Workers}
\label{network}
\end{figure}

\begin{enumerate}

\item\textit{Mixed-Ranked,} containing the largest group of workers with a diverse expertise rate.

\item\textit{High-Ranked,} representing the second largest group of workers, also it contains the largest group of higher experienced rated workers, Yellow and Blue.

\item\textit{Mid-Ranked,} as the third largest community, representing the highest number of average experienced ranked workers, Green. 

\item\textit{Low-Ranked,} presents the smallest community created from mostly lower experienced ranked workers, Gray.

\end{enumerate}





Then, for each community, we measured 4 different network analysis metrics, namely, "Number of Common Neighbors", "Level of Worker Importance", "Closeness Centrality", and "Betweenness Centrality". 
The definition of these metrics are as follows:

   
   
     
\begin{itemize}
   \item \textit{Number of Common Neighbors (${\#CN}$)}:  Measures the number of common workers between any two workers in the sub-network;
   
   \item \textit{Worker Importance (${WI}$)}: Counts the number of incoming weighted edges to a worker to determine a rough estimate of how important the worker is;
   
    \item \textit{Closeness Centrality (${CC}$)}: Measures the average shortest distance from each worker in the community to the other workers;
     
   \item \textit{Betweenness Centrality (${BC}$)}: Measures the extent to which a worker lies on paths between other workers;

\end{itemize}

\subsubsection{Worker Performance} In this study, worker $i$'s performance, in community $k$, is defined as a tuple of  the worker's reliability ($ {RL_{i,k}} $), trustworthiness (${TL_{i,k}}$), and success (${SL_{i,k}}$).
\[
{Worker Performance_{i,k}}  = {({{RL_{i,k}}, {TL_{i,k}}} , {SL_{i,k}})}
\]
It is reported that 38\% of crowd workers provide untrustworthy (i.e. poor quality) submission \cite{eickhoff2011crowdsourcable}\cite{gadiraju2015understanding}. Therefore it is important that not only a worker make a submission but also make a valid submission.
Worker reliability is the probability of receiving a submission by worker $i$, while worker trustworthiness is the probability of receiving a valid submission by worker $i$; worker success represents the probability of a worker making a valid submission for a task and win the task by passing the peer review process.
Note that each worker $i$ in community $k$ is a tuple of the number of registrations ($ {R_{i,k}} $), number of submissions ($ {S_{i,k}} $), and the number of valid submissions ($ {VS_{i,k}} $).
Worker reliability, worker trustworthiness, and worker success are defined mathematically as below:

\textit{Def. 1}:
Worker Reliability Level, ${RL_{i,k}}$, represents the ratio for the
number of tasks registered and submitted by worker i in community $k$, ${S_{i,k}}$, to the total tasks registered by all the workers in the community as:
\begin{equation}
{RL_{i,k}} = \frac{{S_{i,k}}} 
{\sum_{j=1}^{N_{wk}}{R_{j,k}}}
\end{equation}
where $N_{wk}$ is the number of workers in community $k$

Tasks owners trust on receiving not just a submission but a qualified submission after receiving a registration by a worker. Such level of trust is quantified as Worker Trustworthiness Level.

\textit{Def. 2}: Worker Trustworthiness Level, ${TL_{i,k}}$, measures number of valid submissions made by worker $i$ in community $k$ for the tasks the worker registered for, ${VS_{i,k}} $, to the total tasks registered by all the workers in the community as:

\begin{equation}
{TL_{i,k}} = \frac{{VS_{i,k}}} 
{\sum_{j=1}^{N_{wk}}{R_{j,k}}}
\end{equation}

While receiving a valid submission increases a worker's trustworthiness, it does not guarantee the worker's success in task winning. There are different reasons to not win a task even with a valid submission. Therefore, a worker's success level, in this research, is defined as a function of  the worker's winning frequency. 

\textit{Def. 3}: Worker Success Level, ${SL_{i,k}}$, measures the ratio of the total winning frequency for worker $i$ in community $k$, ${W_{i,k}}$, to the total tasks registered by all the workers in the community as:
\begin{equation}
{SL_{i,k}} = \frac{{W_{i,k}}} 
{\sum_{j=1}^{N_{wk}}{R_{j,k}}}
\end{equation}
\subsubsection{Worker Preference}
Worker preference for worker $i$ is defined as a tuple of the worker's efficiency in working on a task that requires  worker's skill set  ($ {EF_{i}} $) and  workers' elasticity involvement with the other workers from the same community in a taken task ($ {EL_{i}} $). 
\[
{Worker Preference_{i}}  = {({EF_{i}},{EL_{i}})}
\]
Worker efficiency, and worker elasticity are defined as follow:

\textit{Def. 5}: Consider $WTech_{i,j}$ as utilization frequency of technology $j$ utilization by worker $i$; for worker $i$, there is a technology-set representing the technologies required to complete tasks that worker $i$ registered for before, $TechSet_{i}$. Worker Efficiency Level, $EF_{i}$, compares utilization frequencies of those technologies by worker $i$ and the total utilization frequency corresponds to all the workers in the worker $i$‘s community, $WrkSet_{i}$, as:  
\begin{equation}
{EF_{i}} = \frac{\sum_{j \in TechSet_{i}}^{}{WTech_{i,j}}} 
{\sum_{j_{1} \in WrkSet_{i}}^{}\sum_{j_{2} \in TechSet_{i}}^{}{WTech_{j_{1},j_{2}}}}
\end{equation}
Each project is divided into some tasks, with $TaskSet_{c}$ representing the set of those tasks. $R_{c,t}$ represents the number of registrants for task $t$ in community $c$. The community Elasticity is defined as the ratio of the total registrants for the set to the maximum registrants of the all communities in a community set, $communitieset$.:

\begin{equation}
{EL_{c}} = \frac{\sum_{j \in TaskSet_{c}}^{}{R_{c,j}}} 
{\max_{c \in communitieset} {\sum_{j \in TaskSet_{c}}^{}{R_{c,j}}}}
\end{equation}

\subsubsection{Worker Strategy}

Worker strategy is defined as tuple of  Workers' comfort, workers' confidence, and  workers' deceitfulness.

\[
{Worker Strategy_{i}}  = {({CT_{i,k}},{CL_{i,k}},{DL_{i,k}})}
\]

Worker comfort, worker confidence, and worker deceitfulness are defined as follow:



\textit{Def. 7}: Workers' Comfort Level, ${CT_{i}}$, measures the ratio of number registrants for a task from lower ranked belt ${LRR_{j}}$ to total task competition level ${TC_{j}}$ on the task $j$ that worker $i$ registered for per worker community.
with $Taskset_{i}$, $LRR_{k,t,i}$ , and $TC_{k,t,i}$ representing
the set of tasks worker $i$ registered for, number of registrations for task $t$ by workers from community $k$ and  ranks lower than with worker $i$' rank, and task competition in terms of the number of total registrations for task $t$ from community $k$

\begin{equation}
{CT_{i,k}} = \frac{\sum_{j \in TaskSet_{i}}^{}{LRR_{k,j,i}}} 
 {\sum_{j \in TaskSet_{i}}^{}{TC_{k,j,i}}}
\end{equation}



\textit{Def. 8}: Workers' Confidence Level ${CL_{i,k}}$ defines as the maximum number of registrations correspond to $Taskset_{i}$, indicating maximum competition level worker $i$ participated in

\begin{equation}
{CL_{i,k}} = \underset{t \in TaskSet_{i}}{\max}\{TC_{k,t,i}\}
\end{equation}
\[
where: {  
    {{S_{i,k}} > 0}}
\]


\textit{Def. 9}: Workes'r deceitfulness Level, ${DL_{i,k}}$, measures the ratio of number of tasks that worker $i$ registered for and did not make any submission, ${ST_{i,k}}$, to total number of tasks that worker $i$ registered for ${R_{i,k}}$.
\begin{equation}
{DL_{i,k}} = \frac{{ST_{i,k}}} 
{{R_{i,k}}}
\end{equation}
\begin{small}
\begin{table*}[!ht]

\caption{Summary of Metrics Definition} 
\centering 
\begin{tabular}{p{2cm} p{4cm} p{10cm}}
\hline
Type & Metrics & Definition \\ 
\hline
\multirow{8}{*}{\parbox{1.5cm}{Workers   Attributes}} 
& \# Registration (R)  & Number of registrants that are willing to compete on total number of tasks in specific period of time.  \\ 
& \# Submissions ($S_{i,t,k}$) & A binary variable correspond to registration $t$ of worker $i$ in community $k$, equal to 1 for a submission and zero for lack of submission  by its submission deadline in specific period of time. \\ 
& \# Valid Submissions ($VS_{i,t,k}$) & a binary variable correspond to registration $t$ of worker $i$ in community $k$, equal to 1 for a submission and zero for lack of a  submission by task's submission deadline and passed the peer review and labeled as either completed or active. \\ 
& Win ($W_{i,t,k}$) & A binary variable correspond to registration $t$ of worker $i$ in community $k$, equal to 1 for a  submission and zero for lack of a  submission  that pass the peer review and labeled as active submission.\\
\hline
\multirow{7}{*}{\parbox{1.5cm}{Workers-Tasks Attributes}}
& Task Status & Completed or failed task. \\
& Task Competition Level (TC) & Total number of workers to register and are willing to compete on a task. \\
& \# Starved Tasks (ST) & Number of Tasks that receives zero submission by its submission deadline and failed. \\
& Technologies (Tech)  & Number of technologies used in task.  \\ 
& Platform (PLT) & Number of platforms used in task. \\ 
& Worker expertise (WTech) & Number of technologies and platforms used in tasks that worker compete on.\\ 

\hline
\label{metrics}
\end{tabular}
\end{table*}
\end{small}

\subsection{Research Questions}  \label{RQ}
Four research questions are formulated as follows:

\begin{itemize}

    \item \textit{RQ1 (Workers Community Patterns)}: How do workers distribute in different communities in a competitive CSD?
    
     This research question aims at providing a general overview of workers’ distribution per identified community based on the members' rank and expertise in the CSD platform; 
    
    \item \textit{RQ2 (Workers' Performance)}: How do different workers' communities impact workers’ performance? Understanding worker reliability, worker trustworthiness, and worker success per identified worker community can be a good measure to indicate worker consistency to perform a task;
    
    \item \textit{RQ3 (Workers' Preferences)}: How do different worker community patterns impact workers’ preferences in taking a task? The degree of worker efficiency and worker elasticity per identified community represent workers’ choice to get involved and compete on tasks;
    
    \item \textit{RQ4 (Workers' Strategy)}: How do workers from different identified communities guarantee their winning in a competition? The degree of worker comfort, worker confidence, and worker deceitfulness per identified community represent workers’ strategy to compete on tasks.

\end{itemize}


\subsection{Dataset}  \label{data}

The dataset from TopCoder contains 403 individual projects, including 4,907 component development tasks (ended up with 4,770 after removing tasks with incomplete information) and 8,108 workers from January 2014 to February 2015 (14 months). The tasks are uploaded as competitions in the platform, where crowd software workers could register and complete the challenges. When the workers submit the final files, they will be reviewed by experts to check the results and grant the scores. This flow creates dynamic attributes of workers that influence successful task delivery in the platform. 

\subsubsection{Data Preparation}

The introduced dataset contains tasks attributes such as required technologies, platform, task description, task status, a monetary prize, days to submit, registration date, and submission date; and workers attributes such as registration date, submission date, valid submission, winning placement, winning status, rating score, and winning score. In this step, workers' performance, preferences, and strategies metrics are not included. \\
We used available data and created attributes such as workers' skillsets (WTech), which are proxied by the number of technologies (\#Tech) required to perform taken tasks by a worker. We created binary variables for each technology (Tech) required in each task, where:
\[
WTech(x,s) =
	\begin{cases}
        1 &\parbox[t]{.35\textwidth}{worker x is an expert in technology s} \\
        0 & \text{otherwise}
    \end{cases}
\]
We created a dataset of workers'  attributes for each identified community. The list and definition of metrics used in the analysis are presented in Table \ref{metrics}.

\begin{small}
\begin{table}[!ht]
\caption{Summary of Different Workers' Belt \cite{saremi2021crowdsim}} 
\centering 
\begin{tabular}{p{2.25cm} p{0.7cm} p{2cm} p{1cm} p{0.7cm}}
\hline

Worker's Rate & Workers' Belt & Rating Range(X) &  Workers\% & p(qualified Sub)\\ 
\hline

\multirow{1}{*}{Lower Rate} 
& Gray &  $ {X < 900} $ & 90.02\% & 0.25 \\
\hline
\multirow{1}{*}{Average Rate} 
& Green & ${900 < X < 1200}$ & 2.88\% & 0.45 \\
\hline
\multirow{2}{*}{Higher Rate}
& Blue & ${1200 < X < 1500}$ & 5.39\% & 0.39 \\
& Yellow & ${1500 < X < 2200}$ & 1.54\% & 0.6\\
\hline
\multirow{1}{*}{Ignored}
& Red & ${X > 2200 }$ & 0.16\% & 0.6 \\
\hline
\label{belts}
\end{tabular}
\end{table}
\end{small}

\textit{Workers' Rate:} TopCoder adopts a numeric worker rating system based on the Elo rating algorithm.  The Elo rating is a method for calculating the relative skill levels of players in competitor-versus-competitor games \cite{saremi2017leveraging}.Based on this numeric rating and a five-level rating scheme, TopCoder divides the worker community into 5 groups. The 5 worker groups are defined into 5 belts of Red, Yellow, Blue, Green, and Gray, which corresponds to the highest skillful workers to the lowest ones\cite{saremi2017leveraging}.The numeric ratings are with respect to three different task categories including algorithm, marathon matches, and development, following sophisticated calculation algorithm. We followed TopCoder rating algorithm to identify the workers belonging to different rating belts in our dataset.Table \ref{belts} summarizes the distribution of workers in different rating belts in our dataset.

\textit{Active Workers:} To identify active workers in the dataset, we sorted workers' activities and labeled workers who had the minimum registration of one as an active worker. This reduced the number of workers from 8180 to 2259. Then, we created the network of active workers based on registration frequency in common tasks.

\subsection{Empirical Study Design}

To empirically investigate the  evaluation framework, we design analysis to answer the four research questions in section \ref{RQ} and conduct experiments using real-world data collected over a period of more than one year, section \ref{data}. Figure \ref{tam} summarizes the steps associated with the study conducted.
In the following subsections, details related to the workers' community pattern, workers' performance analysis, workers' preference analysis and workers strategies analysis are presented.


\subsubsection{Decision Variables}
 In most engineering systems, there are 3 types of variables, namely dependent, independent, and control variables. 
The independent variables are the ones which values are not affected by the other variables in the system. In this research, the independent variables are registration (R), submissions (S), valid Submissions (VS), and win (W).
On the other hand, there are dependent variables, which are extracted from either the independent variables or the other dependent variables. In this research, variables defined in section \ref{model}, i.e equations 1 to 9, such as worker reliability level, worker trustworthiness, worker success level, and worker confidence level, are the dependent variables. The dependent variables are used to provide a tangible understanding toward workers' preferences and performances. 
Also, there is a subset of independent variables named control variables. The control variables are representative of workers' decisions making and directly impact their performances. In this research worker registration (R) and submissions (S) are the control variable. 
\subsubsection{Workers' Community Pattern}

Workers' community pattern helps identify the existing communities of workers among the active workers in our dataset. To identify existing communities, we perform the following: 

first, we analyzed the worker-worker relation by calculating the number of times a pair of workers competed on the same tasks. This analysis provided us with a dataset of "source", "target" and "weight", in which source was the chosen worker, target was the worker who had a common competition history with the source, and weight is the frequency which both the source and target registered for the same tasks. Using this dataset we created the bipartite network of workers.

Second, we identified existing communities in the worker network. To identify the existing communities, we used Clauset-Newman-Moore greedy modularity algorithm \cite{PhysRevE.70.066111} as one of the most used community detection methods. In this step, we found four active communities with a diverse combination of workers.

Third, we analyzed each community by different metrics of the number of common neighbors, worker rank in the network, closeness centrality, and betweenness centrality. Also, we took a deeper look at the worker's diversity in terms of expertise level based on TopCoder definition \cite{saremi2017leveraging}. 

\subsubsection{Workers' Performance Analysis}

We investigate the performance of different ranked workers per community by looking into workers' reliability, trustworthiness, and success. The probability of a worker making a submission after registration for a task will be reported as worker reliability; the probability of a submission passing the peer review and being labeled as valid submission gives the worker's trustworthiness; and, the probability of a submission passing the peer review process and being labeled as win gives the worker's success. In TopCoder, the crowd workers' reliability of competing on the tasks is measured based on the last 15 competitions workers registered for and submitted to. For example, if a worker submitted 14 tasks out of 15 last registered tasks, his reliability is 93\% (14/15). 

\subsubsection{Workers' Preference Analysis}

We investigate the preference of different ranked workers per community to get involved in a task by looking into workers' efficiency and elasticity. The probability of a worker registering for a task with the same technical requirements of workers expertise reports as worker efficiency, and the ratio of the maximum number of registrants per task in a project per community by the maximum number of registrants per task in a project in the platform provides worker elasticity. 

\subsubsection{Workers' Strategy Analysis}

We investigate the potential strategies that higher ranked workers (Blue and Yellow belt) may take per community to assure easier competition level and their success. To that end, we looked into workers’ confidence, the average task competition level that a higher ranked worker makes a submission. Then we looked at the worker's comfort which is the ratio of the number of lower ranked workers register for the task that higher ranked workers registered too. And finally, workers’ deceitfulness is the probability of a higher ranked worker registered for a task and the task starved. The result of this part is relevant since it is an indication of fairness in a competition per identified community of workers.

\section{Empirical Results}

\subsection{Workers community Pattern (RQ1)}

We identified four communities in the network of workers in our dataset. Then, we applied the four most common network analysis metrics on the identified communities to understand the dynamic of workers within them. Based on the result of these analyses, we defined communities as:

\begin{enumerate}

    \item \textit {Mixed Ranked} contains the largest group of workers with a very mixed expertise rate. This community, on average, has 52 common workers per pair of workers. Each worker has a 0.6 probability of being close to the closest neighbor and a 0.2 probability of being in another worker's shortest path. Also, the average worker rank is 0.07, which means there is a 7\% chance that a worker is competing on a task against a highly similar worker;

    \item \textit {High Ranked} represents the second largest group of workers; it contains the largest group of higher rated workers (15\% Blue and  11\% Yellow) among all four communities. Interestingly, workers in this community have the average most significant number of common neighbors (i.e., 88) with a standard deviation of 51. The closeness centrality is 0.59, and the average betweenness centrality is 0.018 with an average worker rank of 0.07;

    \item \textit {Mid Ranked} contains 305 members from the pool of active workers and represents the highest level of average rated workers with 36\% population from Green belt. On average, each worker has 65 common neighbors with average closest centrality of 0.6 and average betweenness centrality of 0.012. Also, the average worker rank is 0.075;
    
    \item \textit {Low Ranked} presents the smallest community, which was also created with mostly lower ranted workers. Workers in this community, on average, have 37 common neighbors with closeness centrality of 0.58 and betweenness centrality of 0.21. Also, each worker, on average, receives a degree of importance of 0.06.
    
\end{enumerate}

Table \ref{netdata} summarized the details of each community.

\begin{small}

\begin{table}[!ht]

\caption{Summary of Identified communities} 
\centering 
\begin{tabular}{p{1.5cm} p{1.5cm} p{0.4cm} p{0.4cm} p{0.4cm}p{0.4cm}p{0.4cm}p{0.4cm}}
\hline

 \multicolumn{2}{c}{Community} & & & {\rotatebox[origin=c]{65}{\#CN}}  & {\rotatebox[origin=c]{65}{WI}} & {\rotatebox[origin=c]{65}{CC}} &{\rotatebox[origin=c]{65}{BC} }\\ 
 \# Workers & Rate Belt & & & &\\
\hline

\parbox[t]{2mm}{\multirow{1}{*}{\rotatebox[origin=c]{60}{Mixed Ranked}{ ---}\rotatebox[origin=c]{90}{906}}}
&Gray: 52\% &\multicolumn{2}{c}{Mean} & 52 & 0.07 & 0.6 & 0.02 \\
&Green: 30\%& & & &\\
&Blue: 11\% & \multicolumn{2}{c}{Std}& 21 &	0.056 & 0.087 &	0.056 \\
&Yellow: 7\% & & & &\\
\hline

\parbox[t]{2mm}{\multirow{1}{*}{\rotatebox[origin=c]{60}{High Ranked}{ --- }\rotatebox[origin=c]{90}{850}}}
&Gray: 59\% &\multicolumn{2}{c}{Mean} & 88	& 0.07 & 0.59 &	0.018\\
&Green: 15\% & & & & \\
&Blue: 15\% &\multicolumn{2}{c}{Std} & 51 &	0.056 &	0.088 &	0.055 \\
& Yellow: 11\% & & & &\\
\hline

\parbox[t]{2mm}{\multirow{1}{*}{\rotatebox[origin=c]{60}{Mid Ranked}{ --- }\rotatebox[origin=c]{90}{305}}}
&Gray: 44\%&\multicolumn{2}{c}{Mean} & 65 & 0.075 &	0.60 &	0.012 \\
& Green: 36\% & & & &\\
&Blue: 12\% &\multicolumn{2}{c}{Std} & 35 &	0.057 &	0.090 &	0.058\\
& Yellow: 8\% & & & &\\
\hline

\parbox[t]{2mm}{\multirow{1}{*}{\rotatebox[origin=c]{60}{Low Ranked}{ --- }\rotatebox[origin=c]{90}{198}}}
& Gray: 72\% &\multicolumn{2}{c}{Mean} & 37 & 0.060 &	0.58 &	0.021 \\
& Green: 11\% & & & &\\
& Blue: 12\% &\multicolumn{2}{c}{Std} & 23 & 0.064 & 0.11 &	0.063 \\
& Yellow: 5\% & & & &\\
\hline
\label{netdata}
\end{tabular}
\end{table}

\end{small}

\textit{Finding 1.1}: High ranked community provides the highest level of interaction among its member.

\textit{Finding 1.2}: Workers in the low-ranked community have the highest probability to register for the same task as any other pair of two workers in the same community.

\subsection{Workers' Performance (RQ2)}

In order to have a better understanding of workers' performance, we studied workers' reliability, trustworthiness, and success per community.  To do so, we created four different datasets based on the identified communities of workers in the network, and then we calculated each of the metrics per dataset.

\textit{Workers' Reliability}: According to Figure \ref{reliability}, while high-ranked community corresponds to an increasing trend of worker reliability as workers' belts increase from 0\% in Gray to 13\% in Yellow.
Interestingly, mixed and mid-ranked communities seem to follow a similar pattern; in both communities, Gray and Blue workers are among the highest reliable workers. While in the mid-ranked community, both Blue and Gray workers provide the same level of reliability of 17\%; in mixed-ranked community, Blue workers provide a slightly higher level of reliability than Gray workers (i.e. 26\% and 24\%). Also, in the mixed-ranked, Green workers are the least reliable workers with a reliability score of 13\%; in the mid-ranked community, Yellow workers bring the lowest level of reliability (i.e. 2\%). 
Moreover, the highest reliability among workers belongs to the Green belt from the low-ranked community with 45\% reliability. The second lowest level of reliability among the low ranked (Blue workers) is equal to the highest level of reliability that mid-ranked workers can provide (i.e. 17\%). And, the second highest reliable workers in the low-ranked community are Yellow belt workers with reliability of 24\%, and the lowest reliability belongs to Gray workers with 13\%.

\begin{small}
\begin{figure}[ht]
\centering
\includegraphics[width=0.75\columnwidth,keepaspectratio]{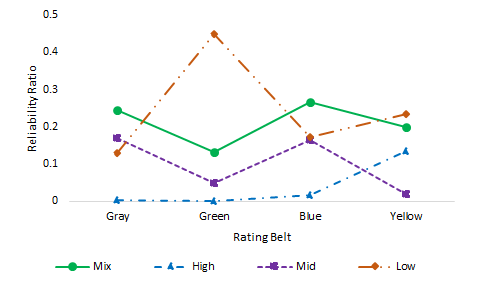}
\caption{Average Worker Reliability per Belt per community of Workers}
\label{reliability}
\end{figure}
\end{small}

We ran a repeated measure one-way ANOVA test on the worker reliability result from the four different worker communities. Based on the ANOVA test results, the worker reliability is significantly different across the existed communities since the p-value is 0.025.

\begin{figure}[ht]
\centering
\includegraphics[width=0.75\columnwidth,keepaspectratio]{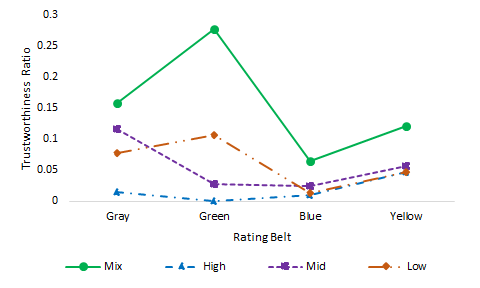}
\caption{Average Workers' Trustworthiness per Belt per community of Workers}
\label{trust}
\end{figure}

\textit{Workers' Trustworthiness}: Besides attracting reliable workers to make a submission, it is important to trust the workers' submissions. Hence, we investigate workers' trustworthiness ratio in returning valid submissions. Figure \ref{trust} represents the distribution of workers' trustworthiness ratio among different workers' belts per community. The mixed ranked community provides the highest trustworthy workers among all the communities. Green workers in this community are the most trustworthy workers with 27\% of trust, followed by Gray and Yellow workers with 15\% and 12\% trust level, respectively; the Blue workers of the mixed-ranked community only provide 6\% of trust.
The high-ranked community contains the least trustworthy workers, while Gray and Green workers in the high-ranked community provide a trust level of 1\% and 0\%; the Blue and Yellow workers are as much trustworthy as their same ranked peers in the low-ranked community (1\% and 4\% respectively).
Also, Gray workers in the mid-ranked community and Green belt workers of the low-ranked community provide 11\% of trust. ANOVA test showed that different communities significantly influence workers' trustworthiness (i.e. p-value is 0.022).

\begin{figure}[ht!]
\centering
\includegraphics[width=0.75\columnwidth,keepaspectratio]{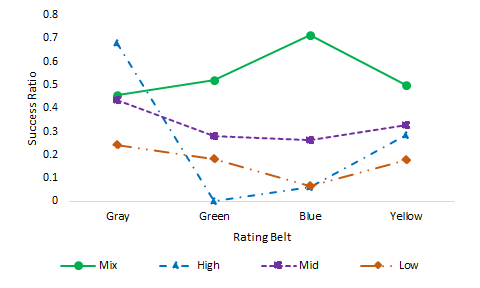}
\caption{Average Workers' Success Per Belt per community of Workers}
\label{success}
\end{figure}

\textit{Workers' Success}:
Workers' performance can be analyzed based on workers' capability of returning a valid submission and win the competition simultaneously. Therefore, we analyzed workers' success based on different workers' ranked belts in various workers' communities.
Figure \ref{success} illustrates the average workers' success per community. As it is shown in Figure \ref{success}, the mixed-ranked community contains the highest successful workers with a success level of 45\%, 51\%, 72\%, and 50\% for Gray, Green, Blue, and Yellow belts, respectively. The second community with the highest success level is mid-ranked, with the highest success rate from the Gray belt with 45\% followed by Yellow, Green and Blue, belts for 32\%, 28\% and 26\%, respectively. Low ranked community is the third place in workers' success level with on average 16\% lower than mid-ranked community. The highest success level in this community belongs to Gray workers for 23\% followed by Green and Yellow for 17\% each.
However, the high-ranked community has the lowest overall success rate; it contains the most successful workers in the pool of workers. Gray belt workers from this community have 67\% success.
ANOVA test showed that worker reliability is significantly different among all four communities with a p-value of 0.047.

\textit{Finding 2.1}: Workers in the mix-ranked community, on average, provide a higher level of trustworthiness and success; however, workers in the low-ranked community are the most reliable ones.

\subsection{Workers' Preference (RQ3)}

For adaptive teams to leverage  CSD  to increase team elasticity, it is critical to understand crowd workers' sensitivity and preference in taking tasks and the rate of team elasticity for them. Therefore, we studied the workers' efficiency and workers' elasticity per different communities of workers. We calculated each of the metrics per community dataset.

\begin{figure}[ht!]
\centering
\includegraphics[width=0.75\columnwidth,keepaspectratio]{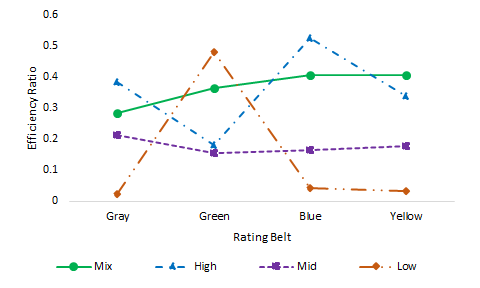}
\caption{Average Workers' Efficiency Per Belt per community of Workers}
\label{efficiency}
\end{figure}

\textit{Worker Efficiency}:
Workers' efficiency can be analyzed based on workers' preferences for performing tasks with requirements from the works' list of expertise. We analyzed workers' efficiency based on different workers ranked belts in different communities of workers. 
Figure \ref{efficiency} displays the average workers' efficiency per community. As it is shown, by increasing workers' rate, in mixed-ranked community, efficiency levels increases. While Gray workers choose to compete on tasks with on average 28\% expertise efficiency, Green workers provide 36\% efficiency, followed by 40\% efficiency for both Blue and Yellow workers. On the other hand, in mid-ranked community, by increasing workers' rate, the efficiency level follows a flat U-shape. Workers in the Gray belt provide the highest level of efficiency in this community with 21\%, followed by 18\% efficiency from Yellow workers. The lowest level of efficiency comes from Green workers with 15\% efficiency and Blue workers with 16\% efficiency.
The average level of efficiency for the entire community puts high-ranked community in second place after mixed-rank community with 35\% efficiency. In this community, Gray and Blue workers have higher efficiency level than mixed ranked (i,e 38\%, and 52\%, respectively) while Green and Yellow workers provide much less efficiency than their peers in mixed-ranked community with 18\% and 34\%, respectively.
In low-ranked group, Gray, Blue, and Yellow workers provide efficiency levels less than 5\%, however, Green workers have on average 48\% efficiency. This puts Green workers from low-raked community in second place after Blue workers from high-ranked community.

ANOVA test showed that worker efficiency is not significantly different among all four communities with a p-value of 0.078.

\begin{figure}[ht!]
\centering
\includegraphics[width=0.75\columnwidth,keepaspectratio]{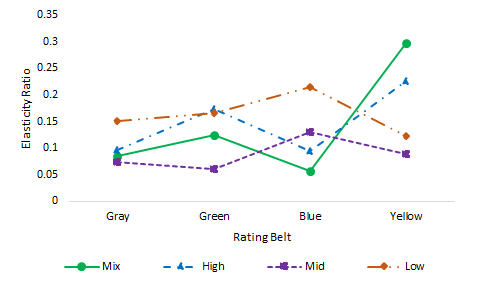}
\caption{Average Workers' Elasticity Per Belt per community of Workers}
\label{elasticity}
\end{figure}

\textit{Workers Elasticity}: After understanding different patterns of workers' efficiency for each community, we investigate the workers' elasticity in crowdsourced tasks per workers rated belt per community. Figure \ref{elasticity} illustrates the distribution of workers' elasticity per worker belt per community. The low-ranked community provides the highest level of worker elasticity with an average of 16\%. In detail, this community of workers has 21\% elasticity for Blue workers, following with 17\% for Green and 15\% for Gray. Yellow workers in this community provide the minimum level of elasticity of 12\%.
Both mixed and high-ranked communities provide almost the same level of elasticity of, on average, 14\%. Interestingly, workers in both communities follow almost the same pattern. In the mixed-ranked community, Yellow workers provide the highest level of 30\% elasticity, followed by Green workers with 12\%, Gray workers with 8\%, and Blue workers with 5\% elasticity. In the high-ranked community, Yellow workers bring the highest level of elasticity in the table for 23\%, followed by Green workers with 17\% elasticity, and both Blue and Gray workers have 10\% worker elasticity.
In the mid-ranked community, Blue workers provide 13\% workers' elasticity; Yellow has 8\%, Gray 7\%, and Green 6\% worker elasticity. These observations provide the mid-ranked community with the least level of elasticity of, on average, 8\%.  
The result of the ANOVA test showed that task workers' elasticity is not significantly different across all four workers' communities (i.e., the p-value is 0.45).

\textit{Finding 3.1}: In general, workers in the mixed-ranked community prefer to work on tasks that they have higher expertise on. Blue workers from the high-ranked community are bringing the maximum workers' efficiency, followed by Green workers in the low-ranked community.

\textit{Finding 3.2}: On average, workers from the low-ranked community provide higher worker elasticity in taking tasks from the same group of tasks.

\subsection{Workers' Strategy (RQ4)}
To understand the strategies that workers with higher rate in each community may take to guarantee their success, or to have an easier competition level per task, we analyzed workers' confidence, workers' comfort, and workers' deceitfulness for the higher rated group of workers per community. Table \ref{Strategy} summarized these metrics.

\begin{small}
\begin{table}[!ht]
\caption{Average Workers’ Strategy per community of Workers } 
\centering 
\begin{tabular}{p{2.25cm} p{1cm} p{1cm} p{1cm} p{1cm}}
\hline
Community & ${R}$ & ${CL}$ & ${CT}$ & ${DL}$\\ 
\hline

\multirow{1}{*}{Mixed Ranked} 
& 12 &  2 & 0.85 & 0.25 \\

\multirow{1}{*}{High Ranked} 
& 11 & 4 & 0.70 & 0.21 \\

\multirow{1}{*}{Mid Ranked} 
& 13& 6 & 0.75 & 0.13\\

\multirow{1}{*}{Low Ranked} 
& 10 & 3 & 0.79 & 0.20\\

\hline
\label{Strategy}
\end{tabular}
\end{table}
\end{small}

\textit{Workers' Confidence}: As it is presented in table \ref{Strategy}, high rank workers from mixed ranked community register for tasks with on average 12 registrants, however, they have the confidence to submit the tasks which received one more registrants beside them (i.e 2). In high ranked community, workers have a confidence level of 3 which means they make a submission when they are competing with another 3 workers. High ranked workers from this community register for tasks with, on average, 11 registrants. Mid ranked community workers register to compete on tasks with, on average, 13 registrants, however, their confidence level of making a submissions is 6 registrants per task . And, low rank community workers have the submissions confidence level of 3, while they register for tasks with average registration of 10.

\textit{Workers' comfort}: According to table \ref{Strategy}, higher-rated workers in the mixed-ranked community try to have the easiest competition with the comfort level of, on average, 0.85. This means these workers register for tasks when, on average, 85\% of registrants are from lower to mid-rated belts (i.e., 10 out of 12). Workers from low-ranked community are in the second rate of easy competition with comfort level of 0.79, following by mid ranked community, which has the comfort level of 0.75. 
Workers from the high-ranked community have a comfort level of 0.7 (8 out of 11 workers are from lower to mid-rated belt). These groups have the most challenging competition level in terms of opponent expertise level among their peers. 

\textit{Workers' Deceitfulness}: Mid-ranked community provides the lowest level of worker deceitfulness, on average 0.13. Low and high-ranked communities are very close, with 0.20 and 0.21 degree of deceitfulness, respectively. Interestingly, the highest level of deceitfulness belongs to the mixed-ranked community with 0.25.

\textit{Finding 4.1}: Higher ranked workers from the mixed-ranked community have not only the lowest confidence in making submissions but also have the highest level of deceitfulness.

\section{Discussion}

\subsection{Workers Community Patterns}

We found four worker community patterns in a network of workers.  
We identified that workers from the high-ranked community have the highest level of interaction among their community members, finding 1.1. Also, we observed that workers in low-ranked community have the highest probability of registering for the same task as any other pair of two workers in the same community, fining 1.2.

\subsection{Workers' Performance}

To successfully crowdsource a software project in a CSD platform, it is essential to understand workers' sensitivity and performance in taking tasks. To that end, this research investigated workers' reliability, workers' trustworthiness, and workers' success per detected worker community. We observed that trustworthiness is higher in a community with more fairness in terms of workers' expertise level (i.e., average rated); however, the highest level of reliability happened in the community with the least experienced workers, finding 2.1.

\subsection{Workers' Preference} 

To ensure that having a successful project in CSD, besides attracting trustworthy and successful workers who make a valid and acceptable submission, it is crucial to attract efficient workers who not only deliver the task but also have the required skillset to make a useful delivery. Also, to understand the reliability of each community of workers in resource shortage per task, we analyzed workers' elasticity level.
And, the results of investigating workers' preference under different workers communities presented that in general, the community of workers with mixed-ranked expertise give higher weight to their skill sets for taking a new task, finding 3.1. Also, the lower-ranked community provides higher worker elasticity and therefore lower risk of resource shortage. However, there is no statistical difference among the communities when analyzing workers' efficiency and elasticity, finding 3.2.

\subsection{Workers' Strategy} 

It is reported that higher rated workers are tempted to apply \textit{"cheap talk"} to soften their competition \cite{archak2010learning}\cite{archak2010money}. To understand how higher-rated workers in different communities may apply such strategies, we investigate workers' confidence level in making a submission, workers' comfort in making sure they have a soft and easy competition, and workers' deceitfulness to check the result of their strategies/approach in task starvation. We observed that workers from the mixed ranked community have a higher potential in applying strategies to assure their success, finding 4.1.


\subsection{Threats to Validity}

First, the study only focuses on competitive CSD tasks on the topcoder platform. Many more platforms exist, and even though the results achieved are based on a comprehensive set of about 5,000 development tasks, the results cannot be claimed externally valid. There is no guarantee the same results would remain the same in other CSD platforms.

Second, many different factors may influence workers' preference, performance, and decision in task selection and completion. Our worker community approach is based on known task-worker attributes in topcoder, and different techniques may lead us to different but almost similar results.

Third, the result is based on the workers -task network only. Workers' communication was not considered in this research. In future, we need to add this level of research to the existing one.

\section{Conclusion and Future Work}

To understand the probability of workers' success and fairness in a crowdsource platform, not only should one understand the available community of active workers in the platform but also need to understand the workers' performance, preference, and chosen strategy in competing on a task. This research investigated the available community of workers by applying the Clauset-Newman-Moore greedy algorithm and observed the workers' behaviors within the identified communities. Then analyzed workers' performance, preferences, and strategies per identified community based on workers rating level in the platform.

Based on statistical analysis, this study can only support that the low-ranked community provides the highest reliability level and mixed-ranked community contains the most trustworthiness workers. 
In future, we would like to evaluate our finding in crowdsourced software development practice and testing the scalability of them in real-time.

\bibliographystyle{ACM-Reference-Format}
\bibliography{mybibfile}

\end{document}